\documentstyle[aps,twocolumn,epsf,floats]{revtex}

\newcommand{\be}{\begin{equation}}
\newcommand{\ee}{\end{equation}}
\newcommand{\bea}{\begin{eqnarray}}
\newcommand{\eea}{\end{eqnarray}}

\newcommand{\la}{\langle}
\newcommand{\ra}{\rangle}

\newcommand{\tr}{\mbox{Tr}}
\newcommand{\rmi}[1]{{\mbox{\scriptsize #1}}}

\begin{document}

\title{Confinement vs. screening in 3d (effective high T) gauge theories}

\author{Owe Philipsen}

\address{Institut f\"ur Theoretische Physik, Philosophenweg 16,
         D-69120 Heidelberg, Germany}


\maketitle

\begin{abstract}
The physics of electric and magnetic screening in non-Abelian 
finite temperature field theory is closely related to the 
non-perturbative properties of the corresponding three-dimensional
effective theories which are confining. I discuss recent lattice results 
obtained for the mass spectrum, string tension and the static potential
in SU(2) pure gauge theory and the SU(2) Higgs model and their 
connection to screening masses. The non-perturbative $O(g_3^2)$ corrections
to the Debye mass are evaluated.
\end{abstract}

\narrowtext


\section{Introduction}
\label{sec1}

Three-dimensional gauge theories play an important role
for high temperature particle physics, since they
constitute the Matsubara zero mode sector of finite temperature quantum field
theories in the imaginary time formalism. In particular, if the
temperatures of interest are larger than the mass scales of the
theory, fermions and non-zero modes may be integrated
out by dimensional reduction \cite{ap} to leave a three-dimensional
effective theory describing all static properties and the equilibrium
thermodynamics of the theory under consideration. While dimensional
reduction is a perturbative procedure, the resulting three-dimensional
gauge theories require non-perturbative techniques in order to 
calculate physical quantities. In particular, in the cases of QCD and
the symmetric phase of the electroweak sector of the Standard Model,
the three-dimensional effective theories are confining and thus
not accessible by perturbative methods. It is then expedient
to split the problem into two parts: perform dimensional
reduction perturbatively, and study the resulting three-dimensional
theory on the lattice. This approach has been particularly successful in
clarifying the order and strength of the electroweak phase transition 
\cite{ew}, and is also applied to the more difficult case
of QCD and the quark gluon plasma, e.g. \cite{qcd}. 
An important question
connected with non-perturbative physics concerns the screening properties
of non-Abelian plasmas like Debye screening, and the much less
understood question of magnetic screening and a corresponding
``magnetic mass" \cite{linde}.

In this contribution, the concept of screening masses is considered
in the framework of three-dimensional gauge theories. 
Starting point is the argument that any static physical quantity
of the finite temperature field theory 
must have a corresponding quantity in the three-dimensional theory, with
which it agrees up to a (perturbative) part 
due to the non-zero Matsubara modes.
In general, a static screening length in finite temperature field theory
is defined as the exponential decay of some {\it spatial} correlation function.
In the dimensionally reduced theory, 
the direction of the correlation may be taken to correspond to (euclidean) time,
and hence the same quantity appears in the spectrum or some other
physical property of the (2+1) dimensional theory. 
In order to fully understand screening in non-Abelian plasmas, it is
therefore necessary to understand the non-perturbative physics
of the underlying three-dimensional theories.
Here I shall discuss the physcial properties of (2+1) dimensional 
gauge theories like the mass spectrum, the static potential, the string
tension, screening of the static potential
and the various correlation functions that serve to 
study them. The connection of these quantities to the
magnetic mass and the Debye mass is discussed. 
Special emphasis is put on the non-perturbative, confining
nature of the theories.
Correspondingly, most of the presented results are drawn from lattice
simulations. 

\section{The SU(2) Higgs model} 
\label{sec2}
As a prominent example corresponding to the dimensionally reduced
electroweak Standard Model \cite{smred}, 
consider the SU(2) Higgs model in (2+1) dimensions,
\bea\label{l3d}
S & =  \int d^3x \; \tr & \left({1\over 2}F_{ij}F_{ij} +
(D_{i}\Phi)^{\dag} D_{i}\Phi \right. \nonumber \\ 
& & \left. + m_3^2 \Phi^{\dag} \Phi
+ 2 \lambda_3 (\Phi^{\dag} \Phi)^2 \right) \, ,
\eea
where all fields are in a $2\times 2$ matrix notation. The gauge 
coupling $g_3$ and the scalar coupling $\lambda_3$ 
have mass dimension 1/2 and 1,
respectively, so the physical properties of the theory are determined
by two dimensionless parameters, $x=\lambda_3/g_3^2$ and $y=m_3^2/g_3^4$.
In the framework of dimensional reduction, these parameters
are of course determined by the four-dimensional couplings $g,\lambda$
and temperature \cite{smred}. In the present context the interest is merely in 
three-dimensional dynamics and this dependence will be supressed.

\begin{figure}\epsfxsize=8cm
\centerline{\epsfbox{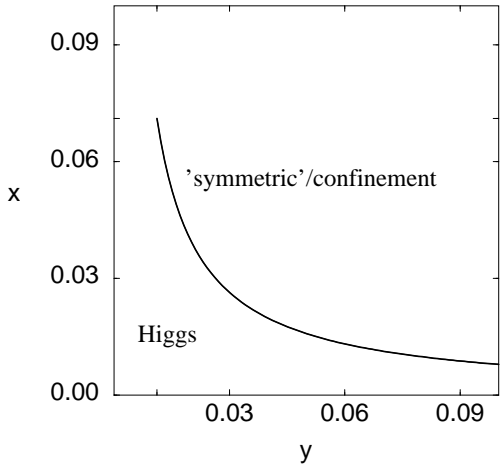}}
\vspace{-3cm}
\caption{Schematic phase diagram of the SU(2) Higgs model with a line
of first order phase transitions.}
\label{pd}
\end{figure}

The phase diagram of the model is schematically shown in Fig.~\ref{pd}.
The perturbative Higgs phase and the non-perturbative symmetric or
confinement phase are for small $x$ separated by a line of first order
phase transitions. With increasing $x$ the transition becomes
weaker until it ends in a point with a second order 
phase transition \cite{ks}. Beyond this point the
tansition disappears entirely to become a smooth crossover, so the phases
are analytically connected. This phase diagram was only established
non-perturbatively in the last few years \cite{bp,kl} and answered
many relevant questions concerning the thermodynamics
of the electroweak phase transition.
Here we are mainly concerned with the non-perturbative properties
of the confinement phase. Due to the connected nature of the phase diagram,
it is possible to smoothly vary the parameters from some point in 
the Higgs region, where the physics is perturbative and well understood,
to the confinement regime and also pure gauge theory, the latter by making
the scalars infinitely heavy so that they decouple.
%
 
\section{Mass spectrum}

All physical properties of the theory are encoded in
gauge-invariant n-point functions. In particular, the mass spectrum
is computed from the exponential fall-off of two-point correlation functions
\be
\lim_{\mid x-y \mid \rightarrow \infty} \la \phi^{\dag}(x)\phi(y)\ra
\sim {\rm e}^{-M\mid x-y \mid},
\ee
where $\phi$ generically denotes some gauge-invariant operator
with quantum numbers $J^{PC}$.
For example, operators with quantum number $0^{++}$ are
\bea \label{op}
R & \sim & \tr \left ( \Phi^{\dag}(x)\Phi(x)\right), \nonumber \\ 
L & \sim & \tr \left( (D_i\Phi)^{\dag}(x) D_{i} \Phi(x) \right),\nonumber \\
P & \sim & \tr \left( F^a_{ij}(x) F^a_{ij}(x)\right),
\eea
where $R$ and $L$ are typically used to measure e.g. the Higgs mass,
whereas $P$ serves to compute 
glueball masses in pure gauge theory.
From each of these operator types analogues with 
quantum numbers other than $0^{++}$ can be constructed
by taking suitable linear combinations with the desired 
rotational and $PC$-symmetry properties. 
A gauge-invariant $1^{--}$ operator, used to compute the mass of
the W-boson, is given by
\be \label{wop}
V^a_{i}\sim \tr \left(\sigma^a \Phi^{\dag}(x) D_{i} \Phi(x) \right).
\ee

A method to compute the mass eigenstates including the spectrum of
higher excitations 
is to measure the full correlation matrix
between all available operators in a given quantum number channel,
$\la \phi^{\dag}_i (x) \phi_j(y) \ra$, where $\phi_i \in \{R,L,P\}$,
and to diagonalise it by a variational calculation to find the
mass eigenstates of the Hamiltonian (for details see \cite{ptw}
and references therein),
\be \label{oleq}
\varphi_i=\sum_k a_{ik} \phi_k.
\ee 
The coefficients $a_{ik}$ are a measure for the overlap or wavefunction
between a given mass eigenstate $\varphi_i$ and the original operators
$\phi_k$.

\begin{figure}\epsfxsize=6cm
\centerline{\epsfbox{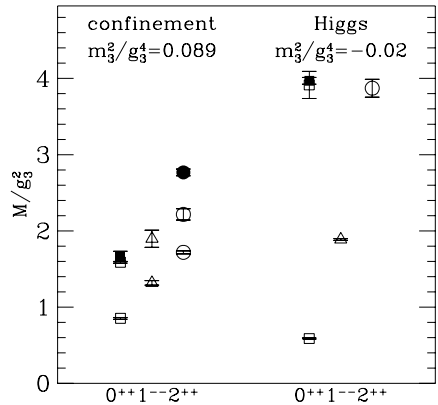}}
\vskip4mm
\caption{ \label{spec} \it
Mass spectrum of the SU(2) Higgs model for $x=\lambda_3/g_3^2=0.0239$ and
two values of $y=m_3^2/g_3^4$, located in the Higgs (right) 
and confinement (left) regions.}
\end{figure}

\begin{figure}\epsfxsize=6cm
\centerline{\epsfbox{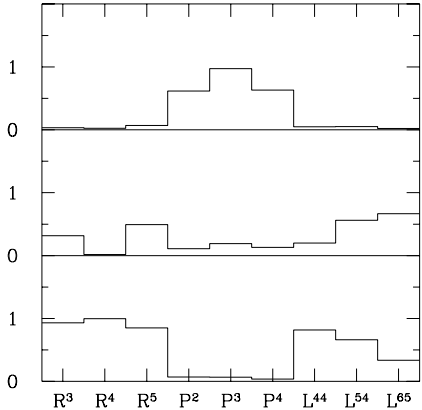}}
\vskip4mm
\caption[]{ \label{ol} \it
The coefficients $a_{ik}$, eq.~(\ref{oleq}), of the operators 
eq.~(\ref{op}) for the lowest three $0^{++}$ states 
in the confinement region (ground state at the bottom). 
Indices on the operators refer
to different smearing levels \cite{ptw}.}
\end{figure}

The results of a lattice calculation \cite{ptw} of the lowest states
of the spectrum with
quantum numbers $0^{++}$, $2^{++}$ and $1^{--}$ at some points in 
the Higgs and confinement region of the phase diagram are displayed
in Fig.\ref{spec}.
In the Higss region we see the familiar and perturbatively 
calculable Higgs and W-boson in the 
$0^{++}$ and $1^{--}$ channels, respectively. There is a large gap 
to the higher excitations which are scattering states.
In the confinement region, in contrast, there is
a dense spectrum of bound states in all channels, very much
resembling the situation in QCD.
Open symbols denote bound states of scalar fields, whereas 
full symbols represent the low lying glueballs.
This identification is based on a 
detailed mixing analysis employing the overlaps (\ref{ol}),
as shown in Fig.~\ref{ol} for the $0^{++}$ channel.
There it is easy to see that the ground state is a bound state of
scalar fields, with almost exclusive contributions from $R,L$-type
operators. The first excitation is a mixture of scalar and gluonic 
contributions. The second excited state, in contrast,
has almost exclusively gluonic content and hence corresponds to a 
glueball.

Furthermore, the numerical values of the glueball 
masses agree within less than 5\%  with those obtained
in lattice simulations of pure SU(2) gauge theory \cite{mike1}.
In the latter, the physical spectrum consists of 
glueballs only. It is very interesting to note that adding scalar matter
fields to the pure gauge theory leaves the glueball spectrum practically
undisturbed, but simply adds an apparently disjoint sector of mesonic
states to the spectrum. This observation is further corroborated by
the fact that the string tension in the confinement region of the
Higgs model is measured to be about 97\% of the one in pure gauge theory
\cite{ptw,mike}.
It has further been established in \cite{ptw} that the glueball masses
as well as the string tension in the confinement phase are insensitive
to variations of the parameters $x,y$, which only influence the masses
of the bound states of scalars. There are preliminary results indicating
that this behaviour is repeated in the SU(2) Higgs model with
scalar fields in the adjoint representation \cite{hp}.

In summary, the mass spectrum in three-dimensional, non-Abelian  
gauge Higgs models consists of a copy of the glueball 
spectrum of the corresponding
pure gauge theory and additional bound states of matter fields.

\section{Magnetic mass vs. physical mass spectrum}

A motivation for the occurrence of the magnetic mass in hot non-Abelian
gauge theories has been given in the contribution of Nair \cite{nair}.
In the following sections the question is considered, whether the magnetic mass
can be identified non-perturbatively in a three-dimensional gauge theory,
and how it might be related to physical quantities of the theory.

In the three-dimensional theory, the magnetic mass according to its
standard definition appears as a pole in the gluon propagator,
\be \label{ww}
\la A^a_i(x)A^b_j(y)\ra \sim {\rm e}^{-m_A \mid x-y \mid}.
\ee
This definition has a caveat. The pole of the gluon propagator
can be shown to be gauge-invariant order by order in (resummed) perturbation
theory \cite{kkr}, but this does not guarantee that the full 
non-perturbative propagator has a pole. On the other hand, since
the gluon propagator itself is gauge-variant,
a fully non-perturbative lattice calculation requires gauge fixing. 
Since that proceeds numerically, gauge-invariance of the mass
is not explicit.
Furthermore, due to the nature of gauge fixing procedures on the lattice,
it is also not immediately evident whether the mass signal extracted really
corresponds to a pole or some other analytic structure.
However, it is important to note that it is quite possible for
a pole in the full gluon propagator to exist, without contradicting
the fact that no asymptotic gluon states can be observed \cite{hoo}. 
A physical asymptotic state is indicated by a pole in the
correlation function of a gauge-invariant operator, and clearly the
gluon propagator does not belong to this class.

The correlator (\ref{ww}) has been evaluated for pure gauge theory
by means of gap equations 
in various different resummation schemes 
at the one loop level \cite{bp,an,jp}, as discussed in \cite{nair}
and summarised in Table \ref{mmag}.
The difference in the results is rather small given that this
is the leading order result for a fully non-perturbative quantity.
A recent two-loop calculation \cite{eb} employing the resummation of \cite{bp}
finds corrections of about 15\% and thus one may hope for 
convergence of the resummed perturbation series.
There are also lattice simulations of the gluon propagator in Landau gauge,
see Table \ref{mmag}. In the simulations in \cite{kar2} different
gauges were employed and no notable gauge dependence was found. 
At present, the reliability and
accuracy of the calculations has not yet reached a fully satisfactory
quantitative level. 
But there is evidence that the gluon propagator
in three dimensions can be evaluated by non-perturbative methods 
in a fixed gauge, and that it produces a mass scale
of the order of $m_A \sim 0.35 - 0.46 g_3^2$.
\begin{table}[t]
\begin{center}
\begin{tabular}{|c|cc|cc|}
               & ref. && $m_A/g_3^2$& \\ \hline
1-loop gap eq. & \cite{an} &   & 0.38 & \\ 
               & \cite{bp,jp}& & 0.28 &\\
               & \cite{cor}  & & 0.25 &\\
2-loop gap eq. & \cite{eb}   & & 0.34 &\\ \hline
lattice        & \cite{kar1} & & 0.35(1)& \\ 
(Landau gauge) & \cite{kar2} & & 0.46(3)& 
\end{tabular}
\end{center}
\caption[]{\label{mmag}
\it Comparison of gluon propagator masses from 
gap equations and lattice simulations.}
\end{table}

Given the difficulties with the propagator pole
the question arises whether the magnetic mass can be identified in
another manifestly gauge-invariant and non-perturbative way. 
If a pole exists in the full gluon propagator,
does it show up in the spectrum or another physical property of the theory? 
Consider again the SU(2) Higgs model discussed above. 
In the Higgs phase, there are
free point like Higgs and W-bosons described by the gauge-invariant
operators $R$ and $V$. The masses extracted from the
correlation functions of these operators agree perfectly 
with those extracted from the poles of the propagators \cite{kar1},
as is expected in a perturbative Higgs regime.
As one moves around the phase diagram to
the confinement phase, the gauge-invariant correlation functions and
the propagators split up. In particular, $m_A$ extracted from
the gluon propagator remains constant, i.e.~it is independent
of the values of the scalar parameters $x,y$ (as it should if it
represents a property of pure gauge theory), whereas
the mass of the $1^{--}$ state rises with $x$ and $y$ to become
much heavier than $m_A$, c.f.~Fig.~\ref{spec}. This is in accord
with its interpretation as a bound state of scalars.
Thus the operator $V$ is not suitable to describe the magnetic mass in
the symmetric phase of the Higgs model, in pure gauge theory 
without scalars it cannot even be defined.
Likewise it is apparent from Fig.~\ref{spec} that 
$m_A$ cannot be identified with any glueball mass (those with quantum
numbers other than $0^{++}, 2^{++}$ are even heavier \cite{mike1}).
In other words, if there is a mass scale associated with
the propagator, it does not show up directly in the spectrum.

\section{A constituent picture}

An indirect way of identifying the propagator mass in the spectrum 
by means of a constituent picture has been proposed in \cite{bp1}.
In the constituent picture, the gauge-invariant composite operators
describe the asymptotic mass eigenstates, whereas the propagator
masses are identified as ``constituent" masses.
In the Higgs phase of the SU(2) Higgs model
the asymptotic states correspond to point particles, i.e. the 
Higgs and W-bosons exist as free particles, 
and hence propagators and gauge-invariant operators
give an equivalent description.
Moving into the confinement phase, the coupling becomes strong 
and what were free particles now become constituents to form bound states.
The latter are the asymptotic states with poles in the correlation 
functions of the composite gauge-invariant operators, whereas
the propagator poles give the constituent masses which do not
correspond to asymptotic states.
This is a unified description where only gauge-invariant operators
correspond to asymptotic states. The fact that asymptotic states 
in the Higgs phase can also be described by the gauge-variant 
propagators is a consequence of the perturbative Higgs nature
of that regime \cite{bp1}.

To the extent that such a constituent picture holds, the 
bound state masses should be the sum of the constituent
masses plus some binding energy.
Counting $\Phi$ as a constituent scalar and $D_{\mu}$ as a 
constituent vector particle, the naive mass formulae for the operators
in eqs. (\ref{op}), (\ref{wop}) are,
\bea
& & m_R \simeq 2 m_{\Phi}\,\quad m_L \simeq 2 m_{\Phi} + 2m_{A}\, \nonumber \\
& & m_P \simeq 4 m_A\, \quad m_V \simeq 2 m_{\Phi} + m_A\, .
\eea
A comparison with the numerical values from lattice simulations
is given in Table \ref{comp}. 
For a test of the constituent picture in the SU(2) adjoint Higgs model,
see \cite{kar2}.
\begin{table}
\begin{center}
\begin{tabular}{|c|cccc|c|}
    &&$J^{PC}=0^{++}$ & & &$J^{PC}=1^{--}$    \\
    &$ R $ &$ L $ & &$ P $ &$ V $  \\ \hline\hline
lattice \cite{ptw}
& 0.839(15) & 1.47(4) & & 1.60(4) & 1.27(6) \\ \hline
const. model & -  & 1.76$\;$ & & 1.67$\;$ & 1.29$\;$
\end{tabular}
\end{center}
\caption[]{\label{comp}
\it Comparison of screening masses from lattice simulations
and a constituent model using $m_A/g_3^2=0.46$, $m_R$
is used to fix the constituent scalar mass.}
\end{table}

Clearly, this naive counting rule works only for operators
that have very good overlap with a definite mass eigenstate.
This is the case for $R,P,V$ but not for $L$, c.f.~Fig.~\ref{ol}.
Since all binding effects have been neglected, such a constituent
picture can at best be qualitatively correct. Moreover, it is not suitable
to describe higher excitations. 
Its purpose in
the present context is not to give precise predictions for the
mass eigenstates, but to furnish
a possible connection between propagator masses and the spectrum of
gauge-invariant operators. 

\section{Static potential and screening}

Another interesting quantity determining the physical properties
of confining theories is the potential energy of static colour sources,
which is calculated from the exponential decay of large Wilson loops,
\be
V(r)=-\lim_{t\to\infty}
\ln\left[\frac{W(r,t)}{W(r,t-1)}\right].
\ee
As in four dimensions, a string of colour flux binds the sources 
(in fundamental as well as adjoint representation)
causing a linear rise in the potential with growing separation $r$.
For sources in the fundamental representation, this linear rise
continues to infinity in pure gauge theory. If fundamental matter
fields are present as in the Higgs model, the string breaks at 
some scale $r_b$, when its energy is large enough to 
produce a pair of scalars. The dynamical scalars are then bound
to the static sources forming a pair of static-light ``mesons", and the
Wilson loop obeys a perimeter law $W(r,t) \sim \exp[-2M(r+t)]$ \cite{bock}
resulting in a saturation of the potential at a constant 
value corresponding to twice the static-light meson energy,
\be \label{asymp}
V(R\rightarrow\infty)=2 M.
\ee
A gauge-invariant two-point function describing the correlation
of such a static-light meson is
\bea
G_\phi(x,y)& = &
 \la \tr \phi^\dagger(t) U^{\rm fund}(x,y)\phi(y)\ra , \nonumber \\
U^{\rm fund}(x,y) & = & {\cal P} \exp({ig\int_x^y\! dx_i A_i^a T^a}).
\label{GPhi}
\eea
At large separations this two-point function falls off with the 
same mass parameter as 
in the Wilson loop perimeter law \cite{bock},
\be
G_\phi(x,y)\sim {\rm e}^{-M\mid x-y \mid}.
\ee

Considering instead the static potential in adjoint representation,
string breaking occurs already in pure gauge theory, because the
adjoint string can couple to gluons that may be pair produced.
The corresponding final state corresponds to a bound state of
the static source and some gluonic constituent and is described by
\bea
G_{F}(x,y)&=&
\la F^a_{ij}(x) U^{\rm adj}_{ab}(x,y)F^b_{kl}(y)\ra , \nonumber \\
U^{\rm adj}_{ab}(x,y) &=&
2 \tr T^a U^{\rm fund}(x,y)T^b [U^{\rm fund}(x,y)]^\dagger ,
\label{GA}
\eea
with now an adjoint Wilson line connecting the field strength tensors.

The correlators eqs.~(\ref{GPhi}) and (\ref{GA}) are manifestly gauge-invariant
because of the string inserted between the fields. If the string is chosen
to be a straight line, then it can be gauged away to be unity in some
axial gauge. In such a gauge and in a perturbative regime 
the correlators coincide with the propagators for the scalar and 
gauge fields. 
Does this connection also hold in a non-perturbative 
regime? If so, the masses of the dynamical fields should be related
to the constituents discussed in the last section.
However, there is one subtlety involved, in that the masses extracted 
from the correlators $G_{\phi},G_{F}$ contain a logarithmic 
divergence due to the self-energy of the Wilson
line representing the static source. Fortunately, due to
superrenormalisability in three dimensions,
the divergence can be computed perturbatively and subtracted \cite{ay,mo},
leaving us with a finite number that is only defined up to a finite
additive renormalisation that has to be fixed by some prescription.
\begin{figure}\epsfxsize=8cm
\centerline{\epsfbox{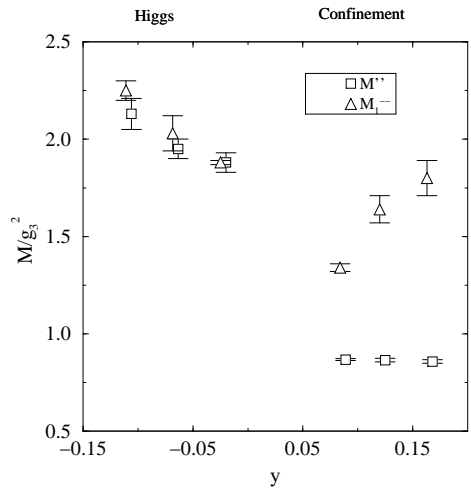}}
\vspace{-3cm}
\caption{The lowest
     physical $1^{--}$ state (the W-boson in the Higgs phase) and
     the mass $M''$ extracted from $G_{F}(x,y)$.
}
\label{giprop}
\end{figure}

Fig.\ref{giprop} shows the mass values extracted from a lattice
simulation of the F-F correlator in comparison with the $1^{--}$ 
bound state mass \cite{mo}. 
In the Higgs phase, the correlator indeed
reproduces the mass of the W-boson as also obtained from the propagator
or the gauge-invariant operator $V$, eq.~(\ref{wop}), after its finite
part has been fixed by matching it to the $V$-correlator at $y=-0.02$.
In the symmetric phase
it stays at a practically constant value of $\sim 0.75 g_3^2$,
which is about half the glueball mass $m_P=1.60(4)g_3^2$ \cite{ptw},
and about twice the propagator mass $m_A=0.46(3) g_3^2$ \cite{kar2}.
Since the field strength tensor contains two covariant
derivatives, $F_{ik}=[D_i,D_k]$, this is again in good agreement with
the naive constituent picture expectation and may be taken as 
another piece of evidence for a gluonic mass unit entering gauge-invariant 
quantities.

\begin{figure}\epsfxsize=9cm
\vspace{-3cm}
\centerline{\epsfbox{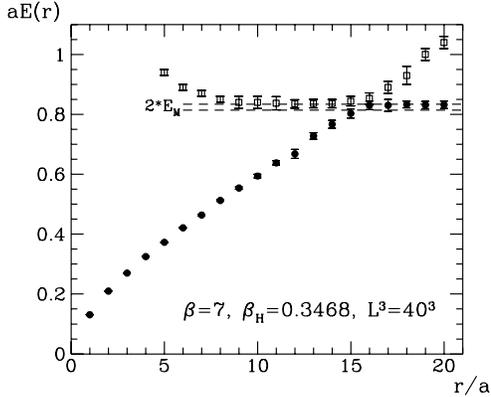}}
\vskip4mm
\caption{\label{break}
The energies of the ground state and the first excited state
for the static potential in fundamental representation. 
The dashed lines indicate
the location of twice the energy of the single meson state, as
extracted from $G_{\phi}$.}
\end{figure}

Finally, another manifestly gauge-invariant and physical quantity
of three-dimensional gauge theories is the string breaking
scale $r_b$ itself. Its size depends on the string tension and the
mass of the dynamical particles that have to be produced to break
the string. The breaking scale for the fundamental representation
potential in the SU(2) Higgs model 
has been calculated in a recent lattice simulation \cite{pw},
by extracting it from the turnover of the potential
as shown in Fig.~\ref{break}. The continuum extrapolation of 
those results gives
\be \label{rb}
r_b g_3^2\approx 8.5,\quad  r^{-1}_b\approx 0.12 g_3^2 \,.
\ee
Again for comparison,
the lightest scalar bound state from Fig.~\ref{spec} is
$m_{R}=0.839(15)g_3^2$, and the lightest glueball $m_{P}=1.60(4)g_3^2$.
A calculation of the adjoint representation potential in pure gauge theory
is still in progress, but
a fairly good estimate for its breaking scale may be obtained by
comparing the final state energy extracted from the F-F correlator
with the energy of the linear part in the potential extrapolated 
from Wilson loop calculations.
The result \cite{mo} is of the same order of magnitude as that
given in eq.(\ref{rb}).

In summary, pure gauge theory as well as the Higgs model contain 
non-perturbative, dynamical mass scales that are much smaller than
any of the masses present in the spectrum. As of yet, it is not understood
what is the origin for those small scales, but intriguing to
speculate whether they may have 
something to do with the previously discussed propagator mass scales.

\section{The Debye mass}

An important concept in the phenomenology of high temperature
QCD is the static electric screening mass, or the Debye mass $m_D$.
Although it has a leading order contribution that is perturbative
\cite{rebhan}, it couples to the three-dimensional 
magnetic sector in next-to-leading
order, and hence requires a non-perturbative treatment as well.
The Debye mass can be expanded as
\be
m_D = m_D^\rmi{LO}+{Ng_3^2\over4\pi}\ln{m_D^\rmi{LO}\over
g_3^2} +
c_N g_3^2 + {\cal O}(g^3T),
\label{md4d}
\ee
where $m_D^\rmi{LO}=(N/3+N_f/6)^{1/2}gT$
and $N_f$ is the number of flavours.
The logarithmic part of the ${\cal O}(g^2)$ correction
can be extracted perturbatively \cite{rebhan},
but $c_N$ and the higher terms are non-perturbative.
To allow for a lattice determination, a non-perturbative
definition was formulated in~\cite{ay}, employing
the SU(N) adjoint Higgs model as
the dimensionally reduced effective theory.
By integrating out the heavy adoint Higgs 
this can be further reduced to the pure SU(N) theory in three dimensions.
The statement of \cite{ay} is that the coefficient $c_N$ entering the Debye 
mass can be determined from the exponential fall-off
of an adjoint Wilson line
with appropriately chosen adjoint charge operators at the ends.
The F-F correlator discussed in the last section is precisely 
such an operator. From its measurement in a lattice simulation
of 3d pure SU(2) in \cite{mo} one finds the complete non-perturbative 
$O(g_3^2)$ corrections to the Debye mass with high precision
to be $c_2=1.06(4)$. Judging from the precision of the result,
the correlator $G_F$ seems to be much more amenable to the analysis
than other operators that have been studied in \cite{lai}.
A calculation for SU(3) is currently in progress.

\section{Conclusions}

Electric and magnetic screening properties of hot non-Abelian
plasmas are non-perturbative phenomena and closely related to 
the physics of confinement in three-dimensional gauge theories.
The physical properties of SU(2) pure gauge theory and the SU(2) Higgs model
in three dimensions have been studied by means of lattice simulations,
and the mass spectrum as well as the static potential 
are known with rather good precision.
The inverse string breaking scales of the potentials in adjoint and
fundamental representation constitute physical mass scales which in
both cases are much smaller than those of the lightest states in the
spectrum. 
It was attempted to establish a connection
between the magnetic mass, as obtained from simulations of the gluon propagator,
and the mass spectrum as well as the static potential 
by means of a constituent picture. Although suggestive, the proposed
connection is not conclusive at this level. In order to make progress
in this direction, it would be necessary to find
a gauge-invariant correlation function giving just one 
gluon constituent mass and thus reproducing the propagator simulations.
Regarding the Debye mass, the full non-perturbative corrections of order 
$\sim g_3^2$ arising from three-dimensional pure gauge dynamics have
been determined following the method of \cite{ay}.

\section{Acknowledgments}

It is a pleasure to thank U. Heinz and 
the organisers for a stimulating conference.

\end{document}